\documentclass[runningheads]{llncs}
\usepackage[hyphens]{url}
\usepackage{graphicx}
\begin{document}

\title{Knowledge Graphs for Digitized Manuscripts in Jagiellonian Digital Library Application}
\titlerunning{Knowledge Graphs for Digitized Manuscripts in JDL}
%
\author{Jan Ignatowicz\orcidID{0009-0005-8716-4175} \and
Krzysztof Kutt\orcidID{0000-0001-5453-9763} \and
Grzegorz J. Nalepa\orcidID{0000-0002-8182-4225}}
\authorrunning{J. Ignatowicz et al.}
%
\institute{Jagiellonian Human-Centered AI Lab, Mark Kac Center for Complex Systems Research, Institute of Applied Computer Science, Faculty of Physics, Astronomy and Applied Computer Science, Jagiellonian University, ul. prof. Stanisława Łojasiewicza 11, 30-348 Krakow, Poland \\
\email{jan.ignatowicz@doctoral.uj.edu.pl, krzysztof.kutt@uj.edu.pl, gjn@gjn.re}}
\maketitle              


\begin{abstract}

Digitizing cultural heritage collections has become crucial for preservation of historical artifacts and enhancing their availability to the wider public. Galleries, libraries, archives and museums (GLAM institutions) are actively digitizing their holdings and creates extensive digital collections. Those collections are often enriched with metadata describing items but not exactly their contents. The Jagiellonian Digital Library, standing as a good example of such an effort, offers datasets accessible through protocols like OAI-PMH~\cite{oai-pmh2002}. Despite these improvements, metadata completeness and standardization continue to pose substantial obstacles, limiting the searchability and potential connections between collections. To deal with these challenges, we explore an integrated methodology of computer vision (CV), artificial intelligence (AI), and semantic web technologies to enrich metadata and construct knowledge graphs for digitized manuscripts and incunabula.

\end{abstract}

\section{Context and Challenges in Digital Cultural Heritage}

The Jagiellonian Digital Library~\cite{jdl}, similar to other institutions, provides digitized collections of manuscripts, incunabula, and other historical artifacts. Although these datasets in majority contain some metadata, their coverage and level of detail differ substantially. Mostly those metadata describe the title, author, creation date, and a few keywords, but they lack detailed information about the physical and visual characteristics of the objects. With having more comprehensive metadata, researchers could explore new and directions, uncovering connections and relationships within broader linked data frameworks.

Additionally, while many organizations provide APIs or open-access protocols like OAI-PMH for data retrieval, others lack such capabilities, requiring manual scraping and extraction. This heterogeneity presents another layer of complexity in building interoperable digital archives.

\section{AI for Metadata Enrichment}

Our approach focuses on enriching existing metadata for digitized manuscripts and incunabula with usage of retrained CV models to find additional information directly from images. These enhanced metadata sets may serve as the basis for knowledge graphs constructions that represent the relationships and attributes of each artifact.

\subsection{Data Preparation and Model Training}

The first step consists of preparing a dataset suitable for training and retraining CV models. Existing widely used pre-trained models, such as YOLOv11~\cite{yolo2024yolo11}, Detectron2~\cite{wu2019detectron2}, DeepLabv3~\cite{chen2017rethinking}, U-Net~\cite{unet}, and HRNet~\cite{wang2020deep}, were tested on selected images from the Jagiellonian Digital Library. However, the unique nature of manuscripts and incunabula, characterized by ornamentations, handwritten texts, and varying degrees of preservation, resulted in these models performing poorly, with detection accuracies falling below acceptable thresholds.

To propose a solution for this problem, we prepared a small dataset of 100 manually annotated images, that captures elements such as text regions, stamps, seals, and decorative features. With the use of this dataset, we retrained the pre-trained models and eventually achieved considerably accuracies in the object detection task. This exercise demonstrates the feasibility of adapting general-purpose CV models to the domain of historical manuscripts. 

\subsection{Metadata Enrichment via Detected Features}

Retrained models were used to analyze digitized manuscripts. They were able to identify key features revealing images' structure and content. The models detected text regions, classified paragraphs and headers, and highlighted decorative elements like ornamentations and illuminated initials. They also located seals, stamps, and marginal notes, what contribute in offering insights into document usage, authorship, and authenticity.

To achieve this, models were trained to recognize specific categories, including paragraphs, stains, stamps, descriptions, signs, signatures, images, ornaments, initials, and headers. Each category reflects a distinct aspect of the manuscripts, from organizing text structure to identifying unique visual elements that vary across creators and historical periods.

\begin{figure}[!ht]
    \centering
    \includegraphics[width=0.85\textwidth]{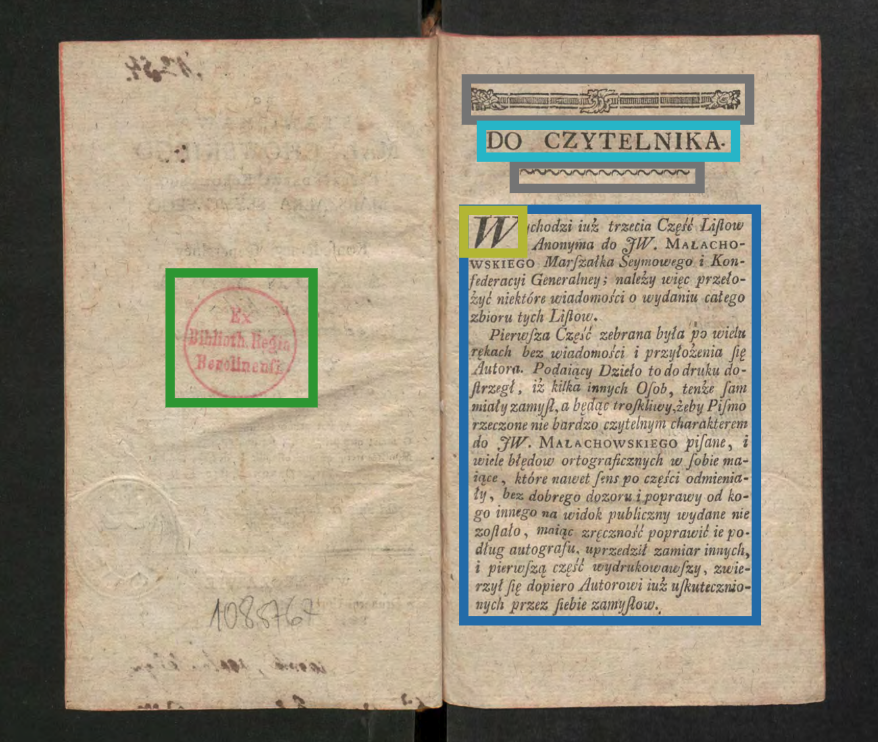}
    \caption{An example image from the created dataset. The image highlights the labeled classes: \textit{stamp}, \textit{initial}, \textit{header}, \textit{ornament}, and \textit{paragraph}.}
    \label{fig:example_dataset_image}
\end{figure}

An example image from the dataset is presented in Figure~\ref{fig:example_dataset_image}.

The detected features were added to the metadata, what enriched each document's description. Text sections are now classified by position, size, and type (e.g., paragraph, header, or signature). Stains and signs inform about a manuscript's condition or its history. Such enhanced metadata enables better search, comparison, and connection between artifacts, what supports deeper research and facilitates integration with broader cultural knowledge networks.

\section{Knowledge Graph Construction}

With enriched metadata, the next step is to build knowledge graphs that semantically represent each artifact’s attributes and relationships. For a manuscript, this step includes linking detected features (e.g., "Stamp X belongs to section Y"). It connects metadata to catalog entries, and integrates with external resources like DBpedia~\cite{auer2007dbpedia} or Wikidata~\cite{vrandevcic2014wikidata}.

Ontologies are used for structuring these graphs. Our approach proposes a modular ontology capturing different types of sttributes. There are physical attributes (dimensions, preservation state), visual features (locations, element classes), and provenance (historical context, authorship, and related works). These connected through Linked Data~\cite{berners2006linkeddata} principles ontologies enable richer exploration, such as tracking recurring seals or the evolution of decorative styles across collections.

\section{Broader Applications and Future Directions}

The proposed methodology supports cultural heritage research by enabling advanced searches for similar decorative styles or stamps. Is also facilitates cross-collection analysis through linked knowledge graphs. Presented methodology supports scalability by allowing larger datasets to be processed as models improve. Future work should focus on expanding the annotated dataset, fine-tuning models for specific artifact types, and automation of ontology generation to increase the methodology’s applicability across diverse collections.

\section{Conclusion}

This work-in-progress highlights an approach to enhance cultural heritage datasets by metadata enrichment. For that this approach combines pretrained computer vision models with semantic knowledge graph construction. By adapting CV models to the unique characteristics of digitized manuscripts and using semantic web technologies, we aim retrieve metadata from images and through that enable richer, more meaningful interactions with digital collections. Although challenges remain, such as expanding datasets and refining ontologies, this interdisciplinary approach proposes advancements in how cultural heritage may be preserved, explored, and understood.

This project underscores the potential of AI and semantic technologies in transforming digital cultural heritage, offering new insights and connections that were previously poorly accessible.

\section*{Acknowledgments}

The research has been supported by a grant from the Priority Research Area (DigiWorld) under the Strategic Programme Excellence Initiative at Jagiellonian University. This publication was funded by a flagship project "CHExRISH: Cultural Heritage Exploration and Retrieval with Intelligent Systems at Jagiellonian University" under the Strategic Programme Excellence Initiative at Jagiellonian University.

%
%

\bibliographystyle{splncs04}
\bibliography{bibliography}

\end{document}